\documentclass[prd,preprint,superscriptaddress,amsmath,amssymb,nofootinbib]{revtex4}
\usepackage{graphicx}
\usepackage{dcolumn}
\usepackage{bm}
\usepackage{amssymb}
\usepackage{amsmath}
\usepackage{epsfig}    
\usepackage{color}
\usepackage{slashed}
\usepackage{hhline}

\def\be{\begin{equation}}
\def\ee{\end{equation}}
\newcommand{\bea}{\begin{eqnarray}}
\newcommand{\eea}{\end{eqnarray}}
\newcommand{\nn}{\nonumber}

\begin{document}

 \begin{flushright} {KIAS-P190xx}, APCTP Pre2019 - 009  \end{flushright}

\title{A radiative neutrino mass model with hidden gauge symmetry \\
inducing semi-annihilating dark matter  }

\author{Takaaki Nomura}
\email{nomura@kias.re.kr}
\affiliation{School of Physics, KIAS, Seoul 02455, Republic of Korea}

\author{Hiroshi Okada}
\email{hiroshi.okada@apctp.org}
\affiliation{Asia Pacific Center for Theoretical Physics, Pohang, Gyeongbuk 790-784, Republic of Korea}

\date{\today}

\begin{abstract}
We  propose two-loop neutrino mass model with gauged hidden $U(1)$ symmetry,
and discuss a Majorana type of dark matter candidate that has semi-annihilation processes in the relic density as well as lepton flavor violations  and muon anomalous magnetic moment.
Also, we demonstrate global analysis to satisfy neutrino oscillation data, lepton flavor violations, and relic density of dark matter candidate and show that semi-annihilation modes play a crucial role in finding observed relic density.
 \end{abstract}
\maketitle

\section{Introduction}
Neutrino mass and dark matter (DM) candidate might be tightly correlated each other, because they possess several similar features such as zero electric charge and weak interactions.
Radiatively induced neutrino mass models are one of the promising candidates not only to explain the neutrino and DM simultaneously but also to correlate them each other~\cite{Ma:2006km}.
{In realizing neutrino mass generation at loop level, we usually require some symmetry to forbid tree level neutrino mass.
Such a symmetry can also stabilize a DM candidate forbidding its decay processes.
A hidden gauge symmetry is one of the attractive candidates to control a neutrino mass model with DM as the SM is described by the gauge symmetry.
}

In this paper, we study a radiative neutrino mass model at loop level with hidden gauged $U(1)_H$ symmetry~\cite{Dey:2019cts, Nomura:2018kdi, Cai:2018upp, Nomura:2018ibs, Nomura:2017wxf, Ko:2017uyb, Ko:2016uft, Ko:2016wce, Ko:2016ala, Ko:2014loa, Ko:2014eqa}, considering fermionic DM candidate with semi-annhilation processes as well as lepton flavor violations.
In our model, Majorana mass term of extra neutral lepton is induced at one-loop level, and then active neutrino mass is generated at one-loop level containing Majorana mass of extra neutral lepton.
Thus, in total, active neutrino mass is generated at two-loop level.
The semi-annihilation processes~\cite{DEramo:2010keq, Cai:2018imb, Belanger:2014bga, Kamada:2017gfc, Cai:2016hne, Cai:2015tam, Cai:2015zza, Aoki:2014cja, Belanger:2012vp} are induced via mixing between neutral components of inert bosons under gauge singlet and triplet which also plays a role in generating nonzero neutrino masses at loop level.
Note that existence of semi-annihilation processes is due to the fact that accidental symmetry in stabilizing DM is $Z_3$ realized by charged assignment for $U(1)_H$.
Then, we numerically show that this semi-annihilation modes are very important to find observed relic density of DM~$\Omega h^2\approx0.12$~\cite{pdg} in our model. Furthermore, we demonstrate the typical order of muon anomalous magnetic moment (muon $g-2$) from our new interactions.

This paper is organized as follows.
In Sec.~II, {we review our model and formulate each of new sector for bosons and fermions including active neutrinos.
Then we discuss neutrino mass generation mechanism,  LFVs,  muon $g-2$, and our DM candidate. As for the DM sector, we formulated the relic density in case where semi-annihilations are added in to the annihilations. Then, we carry out  global analysis numerically to investigate if our model can satisfy neutrino oscillation data, LFVs,
and relic density of DM, and show the allowed regions for several observables. 
In Sec.~III, we devote the summary of our results and the conclusion.}

\section{Model setup and Constraints}
\begin{table}[t!]
\begin{tabular}{|c||c|c|c|c||c|c|c|c|c|c|c|c|c|c|c|c|}\hline\hline  
& ~$L_{L_a}$~& ~$e_{R_a}$~& ~$L'_a$~& ~$E'_a$~& ~$H_1$~& ~$H_2$~& ~$\Delta$~& ~$\varphi_1$~ & ~$\varphi_2$~ &~$\varphi_3$~ &~$\varphi_4$~ & ~$\chi$~  & ~$h_1^+$~   & ~$h_2^+$~ & ~$h_3^+$~  \\\hline\hline 
$SU(2)_L$   & $\bm{2}$  & $\bm{1}$  & $\bm{2}$ & $\bm{1}$   & $\bm{2}$ & $\bm{2}$  & $\bm{3}$  & $\bm{1}$    & $\bm{1}$ & $\bm{1}$ & $\bm{1}$  & $\bm{1}$ & $\bm{1}$    & $\bm{1}$ & $\bm{1}$   \\\hline 
$U(1)_Y$    & -$\frac12$  & -$1$ & {-$\frac12$} & $-1$ & $\frac12$  & $\frac12$   &{$1$} &{$0$}  &{$0$}   &{$0$}  &{$0$}  &{$0$} & $1$ & $1$ & $1$ \\\hline
$U(1)_H$    & $0$  & $0$ & {$-x$} & $-4x$  & $0$  & $3x$  &{$2x$}  &{$3x$} & $(\kappa - 3)x$ & $(11- 2 \kappa)x$ & $(\kappa - 5)x$  &{$x$} & $2x$ & $\kappa x$ & $(2 \kappa -6) x$  \\\hline
\end{tabular}
\caption{Charge assignments of the our lepton and scalar fields
under $SU(2)_L\times U(1)_Y\times U(1)_H$, where $x \neq 0$, the loser index $a$ is the number of family that runs over 1-3,
all of them are singlet under $SU(3)_C$, $\kappa$ is free parameter for charge assignment, and $\Delta$ and $\chi$ are expected to be inert bosons. }\label{tab:1}
\end{table}

In this section we formulate our model.
{As for the fermion sector, we introduce three families of vector fermions $L'$ and $E'$ with respectively $(2,{-1/2},-x)$ and $(1,-1,-4x)$ charges under the $SU(2)_L\times U(1)_Y\times U(1)_H$ gauge symmetry, where $ x\neq0$.
As for the scalar sector, we add an $SU(2)_L$ doublet $H_2$ with $(1/2,x)$,
 triplet $\Delta$ with $(1,2x)$, two charged scalar $h_{\alpha = 1,2,3}^+$ with $\{ (1,2x), (1,\kappa x), (1, (2\kappa -6)x) \}$, and three singlets $\varphi_{A=1,2,3,4}$  with {$\{ (0,3x), (0, (\kappa-3)x), (0, (11-2 \kappa)x ), (0,(\kappa-5)x) \}$} and  $\chi$ with {$(0,x)$} charges under the $U(1)_Y\times U(1)_H$ gauge symmetry in addition to the SM-like Higgs that is denoted as $H_1$, where  $\Delta$ and $\chi$ are expected to be inert scalar fields. 
The parameter $\kappa$ appearing in charge of scalar fields is free parameter and it should not be some specific values to spoil inert condition for $\chi$ and $\Delta$; 
for example $\kappa=2$ allows $\chi \varphi_2^*$ term endangering inert condition for $\chi$ not to develop its VEV.
Note also that we introduce three $SU(2)_L$ charged scalar fields which are required to accommodate one-loop generation of Majorana mass term of neutral component in $L'$ with inert condition for $\chi$;
we will explain more detain in Sec.II.A.
 }
Here we respectively write the nonzero vacuum expectation values {(VEVs)} of $H_1$, $H_2$, and $\varphi_A$ by $\langle H_1\rangle\equiv v_H/\sqrt2$, $\langle H_2\rangle\equiv v_{H'}/\sqrt2$ and $\langle \varphi_A \rangle \equiv v_{\varphi_A}/\sqrt2$ after the spontaneous electroweak symmetry breaking.
Relevant field contents and their assignments are summarized in Table~\ref{tab:1}, where the quark sector is the same as the SM and it is omitted from the table.
{The scalar fields are written by their components as follows
\begin{align}
& H_{1} = \begin{pmatrix} \phi_{1}^+ \\ \frac{1}{\sqrt{2}} (v_H + h_1 + i a_1) \end{pmatrix}, \quad H_{2} = \begin{pmatrix} \phi_{2}^+ \\ \frac{1}{\sqrt{2}} (v_{H'} + h_2 + i a_2) \end{pmatrix}, \nonumber \\
& \Delta = \begin{pmatrix} \frac{\delta^+}{\sqrt{2}} & \delta^{++} \\ \frac{1}{\sqrt{2}} (v_\Delta+\delta_R+ i \delta_I) & - \frac{\delta^+}{\sqrt{2}} \end{pmatrix}, \quad
\varphi_A = \frac{1}{\sqrt{2}} (v_{\varphi_A} + \varphi_{R_A} + i \varphi_{I_A}), \quad \chi = \frac{1}{\sqrt{2}} (\chi_R + i \chi_I),
\end{align}
where $A=1-4$ distinguish singlet scalar fields generating VEVs.
One linear combination of singly charged components becomes Nambu-Goldstone(NG) boson which is absorbed by $W$ boson and 
two degrees of freedom in CP-odd scalars will be absorbed by $Z$ and $Z'$ bosons as neutral NG bosons.
}
The renormalizable Yukawa Lagrangian under these symmetries is given by
\begin{align}
-{\cal L_\ell}
& =  y_{\ell_{aa}} \bar L_{L_a} H_1 e_{R_a}  +  f_{ab}\bar L_{L_a} L'_{R_b}\chi  
 +  g_{L_{ab}} \bar L'^c_{L_a}(i\tau_2) \Delta L'_{L_b}
 +  g_{R_{ab}} \bar L'^c_{R_a}(i\tau_2) \Delta L'_{R_b} \nn\\
&+M_{L_{aa}} \bar L'_{L_a} L'_{R_a} + {M_{E_{aa}} \bar E'_{L_a} E'_{R_a} }
 {+ y_{E} \bar L' E' H_2 + y_{L'} \bar L'^c L' h_1^+ + {\rm h.c.}}, \label{Eq:yuk}
\end{align}
where lower indices $(a,b)=1$-$3$ are the number of families, $\tau_2$ is the second Pauli matrix, $y_\ell$ and either of $g_{L/R}$ or $M_L, {M_E}$ can be diagonal matrix without loss of generality, and we omitted flavor index for the last two terms. Here, we assume both $g_{L/R}$ and $M_L, {M_E}$ to be diagonal for simplicity.
The charged-lepton mass matrix is then given by $m_\ell=y_\ell v_H/\sqrt2$. 

{
\noindent \underline{\it $Z'$ boson from $U(1)_H$}: 
After spontaneous symmetry breaking of $U(1)_H$, we have massive $Z'$ boson.
The mass of $Z'$ is given by
\begin{equation}
m_{Z'}^2 = 9x^2 v_{H'}^2 + \sum_{A=1}^4 (Q^H_A)^2 v_{\varphi_A}^2,
\end{equation}
where $Q^H_A$ denote $U(1)_H$ charge of $\varphi_A$.
This $Z'$ couples to the SM fermions only through kinetic mixing and/or $Z$-$Z'$ mixing since it is associated with hidden gauge symmetry.
In this paper we assume $Z'$ is much heavier than $Z$ boson and the mixing is small.
We thus do not discuss $Z'$ effect further in the following analysis.
}

\noindent \underline{\it Scalar potential and VEVs}:
The  scalar potential in our model is given by 
\begin{align}
{\cal V} = &  
 \mu_\chi^2 |\chi|^2+ \mu_1^2 |H_1|^2  + \mu_2^2 |H_2|^2 + \mu_\Delta^2 {\rm Tr}[\Delta^\dag\Delta] { +  \sum_{A=1}^4 \mu_{\varphi_A}^2 |\varphi_A|^2 
   +  \sum_{\alpha=1}^3 \mu_{h^\pm_\alpha}^2 h^+_\alpha h^-_\alpha } \nn \\
& { + \mu_1 (H_2^\dagger H_1 \varphi_1 + h.c.) + \lambda_1 (h_3^+ h_2^- \varphi_1 \varphi_2^* + h.c.) + \lambda_2 (h_3^+ h_1^- \varphi_1^* \varphi_3 + h.c.) }
\nn\\
& {  + \lambda_3 (\chi^2 \varphi_2^* \varphi_4 +h.c.) + \lambda_\chi (\chi^*  \varphi_1 \varphi_2^*\varphi_4 + h.c.) + \tilde{\lambda}_\chi (\chi^* \varphi_3 \varphi_4^2  + h.c.)} \nn \\
&+ \lambda(H_1^T(i\tau_2)\Delta^\dag H_2\chi^* +{\rm h.c.})  {+ \lambda' (H_1^T (i \tau_2) H_2 h_2^- \varphi_2 +{\rm h.c.})  + \lambda'' (\varphi_2^2 \varphi_3 \varphi_4 + h.c.) } \nn\\
&  + \lambda_{\chi}|\chi|^4 + \lambda_{H_1}|H_1|^4+ \lambda_{H_2}|H_2|^4
+ \lambda_{\Delta}{\rm Tr}[|\Delta^\dag\Delta|^2]+ \lambda'_{\Delta}({\rm Tr}[\Delta^\dag\Delta])^2\nn\\
& {+ \sum_{A=1}^4 \Big[ \lambda_{\varphi_A }|\varphi_A |^4 +\lambda_{\varphi_A \chi}|\varphi_A|^2|\chi|^2+\lambda_{\varphi_A H_1}|\varphi_A|^2|H_1|^2+\lambda_{\varphi_A H_2}|\varphi_A|^2|H_2|^2 } \nn \\
& { +\lambda_{\varphi_A \Delta}|\varphi|^2 {\rm Tr}[\Delta^\dag\Delta] + \lambda_{Ah_1^+} |\varphi_A|^2 h_1^+ h_1^- + \lambda_{Ah_2^+} |\varphi_A|^2 h_2^+ h_2^-
+ \lambda_{Ah_3^+} |\varphi_A|^2 h_3^+ h_3^-  \Big] } \nn\\
&+\lambda_{\chi H_1}|\chi|^2|H_1|^2+\lambda_{\chi H_2}|\chi|^2|H_2|^2
+\lambda_{\chi \Delta}|\chi|^2 {\rm Tr}[\Delta^\dag\Delta] {  + \sum_{\alpha = 1}^3 \lambda_{\chi h_\alpha^+} |\chi|^2 h_\alpha^+ h_\alpha^-} \nn\\
&+\lambda_{H_1H_2}|H_1|^2|H_2|^2+\lambda'_{H_1H_2}|H_1^\dag H_2|^2+\lambda_{H_1 \Delta}|H_1|^2{\rm Tr}[\Delta^\dag\Delta] 
+\lambda'_{H_1 \Delta}\sum_{i=1}^3(H_1^\dag \tau_i H_1){\rm Tr}[\Delta^\dag\tau_i\Delta] \nn\\
&+\lambda_{H_2 \Delta}|H_2|^2{\rm Tr}[\Delta^\dag\Delta] 
+\lambda'_{H_2 \Delta}\sum_{i=1}^3(H_2^\dag \tau_i H_2){\rm Tr}[\Delta^\dag\tau_i\Delta]   \nn \\
& {+ \sum_{\alpha =1}^3 \lambda_{\Delta h_\alpha^+}  {\rm Tr}[\Delta^\dagger \Delta] h_\alpha^+ h_\alpha^- + \sum_{j=1,2} \sum_{\alpha=1}^3  \lambda_{H_j h_k^+}  |H_j|^2 h_\alpha^+ h_\alpha^- },
\label{Eq:potential}
\end{align}
where $\tau_i$(i=1-3) is Pauli matrices. 
{In realizing neutrino mass generation at loop level, we require $\chi$ and $\Delta$ not to develop its VEV.
Since three are terms proportional to $\chi$, we have non-trivial relation among parameters and the other VEVs to satisfy $\partial V/\partial \chi =0$ with $\langle \chi \rangle =0$ such that
\begin{equation}
\lambda_\chi v_{\varphi_1} v_{\varphi_2} v_{\varphi_4} + \tilde \lambda_\chi v_{\varphi_3} v_{\varphi_4}^2 = 0.
\label{eq:chiVEVcond}
\end{equation} 
The other non-zero VEVs can be derived by solving the conditions $\partial V/\partial v_{H,H', \varphi_A} =0$ for the scalar potential.
The term with $\lambda_3$ coupling lead mass difference between $\chi_R$ and $\chi_I$ which is necessary to obtain non-zero contribution to active neutrino mass as we see below.
Here we assume singlet scalar VEVs are much higher than electroweak scale, and we obtain potential for two Higgs doublet sector after $\varphi_A$ developing VEVs.
The VEVs of two Higgs doublet are then obtained from these equations; 
\begin{align}
&  \left(\mu_1^2 + \sum_{A=1}^4 \lambda_{H_1 \varphi_A} v_{\varphi_A}^2  \right)v_{H} + \frac{\mu}{2\sqrt{2}} v_{\varphi_1} v_{H'}  + \lambda_{H_1} v_{H}^3 + \frac{\lambda_{H_1 H_2}}{2} v_{H} v_{H'}^2 + \frac{\lambda'_{H_1 H_2}}{2} v_{H} v_{H'}^2 = 0, \\
&  \left(\mu_1^2 + \sum_{A=1}^4 \lambda_{H_1 \varphi_A} v_{\varphi_A}^2  \right)v_{H'} + \frac{\mu}{2\sqrt{2}} v_{\varphi_1} v_{H} + \lambda_{H_2} v_{H'}^3 + \frac{\lambda_{H_1 H_2}}{2} v_{H'} v_{H}^2 + \frac{\lambda'_{H_1 H_2}}{2} v_{H'} v_{H}^2  = 0, 
\end{align}
Note also that the term with $\mu$ in the potential is necessary to avoid massless Goldstone boson from two Higgs doublet sector.
Inert singlet $\chi$ and neutral component of inert triplet $\Delta$ can mix through the term proportional to $\lambda$ in the potential.
The mass matrix for $(\chi_{R/I}, \delta_{R/I})$ is given by 
\begin{equation}
\begin{pmatrix} M_{\chi_{R/I}}^2 & \frac{\lambda}{\sqrt{2}} v_H v_{H'} \\  \frac{\lambda}{\sqrt{2}} v_H v_{H'} & M_{\delta_{R/I}}^2 \end{pmatrix},
\end{equation}
where $M^2_{\chi_{R/I}}$ and $M^2_{\delta_{R/I}}$ are mass squared parameters for $\chi_{R/I}$ and $\delta_{R/I}$ obtained from the scalar potential.
Here, we define the mixing between each of the neutral component for $\chi$ and $\Delta$
as~\cite{Nomura:2019jxj} 
\begin{align}
&\chi_{R}=c_{R} H_1 + s_R H_2,\ 
\delta_{R}=-s_{R} H_1 + c_R H_2,\\
&\chi_{I}=c_{I} A_1 + s_I A_2,\ 
\delta_{I}=-s_{I} A_1 + c_I A_2,\\
& s_{2R}= \frac{\sqrt{2} \lambda v_H v_{H'}}{m_{H_2}^2-m^2_{H_1}}, \
s_{2I}= \frac{\sqrt{2} \lambda v_Hv_{H'}}{m_{A_2}^2-m^2_{A_1}}, 
\end{align}
where $s_{R/I}$ is short-hand notation for $\sin\theta_{R/I}$.

The two Higgs doublet sector is similar to type-I two Higgs doublet model in which only one Higgs doublet couples to the SM quarks and leptons.
Thus we omit analysis of the sector since its nature is already well studied.  
}

\begin{figure}[tb]
\begin{center}
\includegraphics[width=10.0cm]{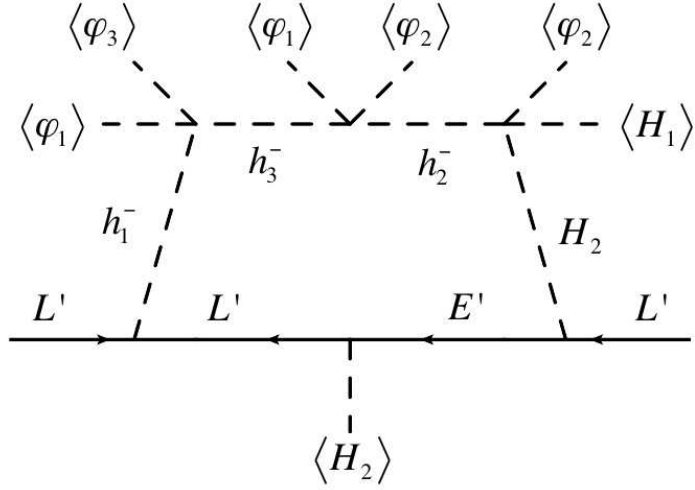}
\caption{Feynman diagram to generate the masses of $\mu_{L/R}$.}
\label{fig:mu_mass}
\end{center}\end{figure}

\subsection{Neutral fermion masses}
\noindent \underline{\it Heavier neutral sector}:
{Firstly Majorana masses of extra neutral fermion are generated at one-loop level by the diagram in Fig.~\ref{fig:mu_mass}. 
The Majorana masses are given by parameters $y_E$, $y_{L'}$, charged scalar masses and mixing among charged scalar bosons.
Here, for simplicity, we simply parametrize Majorana mass of neutral components of $L'_{L(R)}$ as $\mu_{L(R)}$.
Before proceeding we explain necessity of three charged scalar $h^+_{\alpha}$. 
The effective operator generating Majorana mass of $L'$ is written as $(H_2^\dagger H_1)(L'L') \times (\text{product of singlet scalars})$ in our scenario.
Since $(H_2^\dagger H_1)(L'L')$ part has $U(1)_H$ charge $x$ the remaining part should have charge $-x$. 
Thus if  the "product of singlet scalars" does not contain number of scalar less than 4 we obtain a renormalizable term proportional to $\chi$ in the potential.  
To avoid the situation, we need at least three singly charged scalars.
}

Then extra neutral fermion mass matrix in basis of $(N_R,N^C_L)^T$ is given by~\cite{Kajiyama:2012xg, Nomura:2016vxr, Catano:2012kw}
\begin{align}
M_N
&=
\left[\begin{array}{cc}
\mu_R & M_L^T   \\ 
M_L &  \mu_L \\ 
\end{array}\right],
\end{align}
where $\mu_{L/R}\equiv \frac{g_{L/R} v_\Delta}{\sqrt2}$.
Since  $\mu_{L/R}<<M_L$, the mixing can be maximal.
Thus, we formulate the eigenstates in terms of the flavor eigenstate as follows:
\begin{align}
 M_\pm \simeq M_L\pm \frac{\mu_L+\mu_R}{2},\quad
 \left[\begin{array}{c}
N_R   \\ 
N^C_L  \\ 
\end{array}\right]
=\frac{1}{\sqrt2}
\left[\begin{array}{cc}
i_{3\times3} & 1_{3\times3}   \\ 
-i_{3\times3} & 1_{3\times3} \\ 
\end{array}\right]
\left[\begin{array}{c}
\psi_{R_1}   \\ 
\psi_{R_2}  \\ 
\end{array}\right],
\end{align}
where we symbolically denote the mixing to be $\Omega$ in the righthand side of the above equation and $[\psi_{R_1},\psi_{R_2}]^T$ represents the mass eigenstate.
{The heavy singly-charged fermions $E'$ and $e'$ in $L'$ also mix each other through $y_E$.
However since this sector does not affect our phenomenology, we simply assume $y_E v_{H'}/\sqrt{2}<< M_E,M_L$ so that we work on $e'$ or $E'$ as mass eigenstates.}

\noindent \underline{\it  Active neutrino sector} :
The active neutrino mass matrix is induced at two-loop level~\footnote{Similar types of diagrams are found in refs.~\cite{Kajiyama:2013zla, Kajiyama:2013rla, Nomura:2016pgg}.}, and it is formulated by
\begin{align}
m_\nu &= 2\sum_{\alpha=1}^6 \frac{Y_{ia} D_{N_a} Y^T_{a j} }{(4\pi)^2}
\left(3 {\cal M}_1^2 F_1^a - {\cal M}_2^4 F_2^a + {\cal M}_3^6 F_3^a \right),\\
&{\cal M}_1^2=c_R^2 m^2_{H_1} + s_R^2 m^2_{H_2} -c_I^2 m^2_{A_1}-s_I^2 m^2_{A_2},\\
&{\cal M}_2^4=(s_R^2-s_I^2) m^2_{H_1}m^2_{A_1} + (c_R^2-c_I^2) m^2_{H_2}m^2_{A_2} \nn\\
 &\hspace{1cm}-(s_I^2 m_{H_2}^2 - s_R^2 m_{A_2}^2)(m^2_{H_1}+m^2_{A_1})
-(c_I^2 m_{H_1}^2 - c_R^2 m_{A_1}^2)(m^2_{H_2}+m^2_{A_2}),\\
&{\cal M}_3^6 =(c_I^2 m_{H_1}^2 - c_R^2 m^2_{A_1}) m_{H_2}^2 m_{A_2}^2 
+(s_I^2 m_{H_2}^2 - s_R^2 m^2_{A_2}) m_{H_1}^2 m_{A_1}^2,\\
F_\rho^a&=\int_0^1
\frac{\Pi_{k=1-5}dx_k\delta(1-\sum_{k=1-5}x_k)}{[x_1 D_{N_a}^2+ x_2 m_{H_1}^2+ x_3 m_{A_1}^2+ x_4 m_{H_2}^2+ x_5 m_{A_2}^2]^\rho},
\ {\rm \rho=1,2,3},
\end{align}
where $D_N\equiv{\rm diag}(M_-,M_+)$, and $Y_{ib}\equiv \sum_{a=1-3}f_{ia}\Omega_{ab}/\sqrt2$. 
Neutrino mass eigenvalues ($D_\nu$) are given by $D_\nu=U^T m_\nu U$, where $U$ is  the observed lepton mixing matrix.
Once we define $m_{\nu} \equiv Y {\cal M} Y^T$, one can rewrite $Y(f)$ in terms of the other parameters~\cite{Casas:2001sr, Chiang:2017tai} as follows:
\begin{align}
f_{ik}=\sqrt2 \sum_{\alpha=1}^6 U^T_{ij} \sqrt{D_{\nu_{jj}}} O_{j\alpha} {\cal M}_{\alpha\alpha}^{-1/2} \Omega^\dag_{\alpha k},
\end{align}
where $i,j,k=1-3$, $\alpha=1-6$, and $O$ is a three by six arbitrary matrix with complex values, satisfying $OO^T=1_{3\times3}$ ($O^TO\neq1_{6\times6}$), and $|f|\lesssim \sqrt{4\pi}$ is imposed not to exceed the perturbative limit.

\begin{figure}[t]
\begin{center}
\includegraphics[width=10cm]{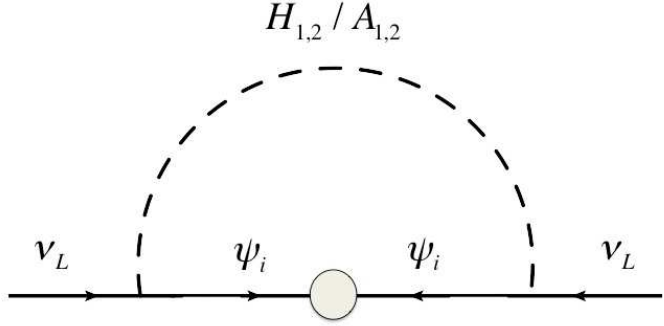}
\caption{The diagram inducing active neutrino mass where the gray circle indicate diagram in Fig.~\ref{fig:mu_mass}.}   \label{fig:diagram}
\end{center}\end{figure}

\noindent \underline{\it  Lepton flavor violations(LFVs) and Muon $g-2$:}
$\ell_i \to \ell_j \gamma$ arise from the term $f$ at one-loop level, and its form can be given by~\cite{Lindner:2016bgg, Baek:2016kud}
\begin{align}
 {\rm BR}(\ell_i\to\ell_j\gamma)= \frac{48\pi^3\alpha_{\rm em} C_{ij} }{{\rm G_F^2}}\left(1+\frac{ m_{\ell_j}^2}{ m_{\ell_i}^2}\right)|{\cal A}_{ij}|^2,
 \end{align}
where $C_{21}=1$, $C_{31}=0.1784$, $C_{32}=0.1736$, $\alpha_{em}$ is the electromagnetic fine structure constant, $G_F=1.166\times10^{-5}$ GeV$^{-2}$, and
 \begin{align}
{\cal A}_{ij} &=\sum_{a=1}^3
 \frac{f_{ja}  f^\dag_{a i}} {64\pi^2}
\sum_{J=H_{1,2},A_{1,2}}{\cal S}_J
\left[\frac{2+3r_{aJ}-6r^2_{aJ}+3r^3_{aJ}+6r_{aJ}\ln r_{aJ}}{6m^2_{J}(1-r_{aJ})^4}\right],
\end{align}
and $r_{aJ}\equiv\frac{M^2_{L_a}}{m^2_J}$, $S_{H_1}\equiv c_R^2, S_{H_2}\equiv s_R^2, S_{A_1}\equiv c_I^2, S_{A_2}\equiv s_I^2$.
These BRs are constrained by the experimental upper bounds given by~\cite{TheMEG:2016wtm, Aubert:2009ag,Renga:2018fpd,Lindner:2016bgg}
\begin{align}
{\rm BR}(\mu\to e\gamma)\lesssim 4.2\times10^{-13},\quad 
{\rm BR}(\tau\to e\gamma)\lesssim 3.3\times10^{-8},\quad
{\rm BR}(\tau\to\mu\gamma)\lesssim 4.4\times10^{-8},\label{eq:lfvs-cond}
\end{align}
which will be imposed in our global analysis with numerical calculation.

We obtain new contribution to the muon $g-2$ ($\Delta a_\mu$) from the same diagrams for LFVs and it is found as
\begin{align}
&\Delta a_\mu \approx - 2 m_\mu{\cal A}_{22},\label{eq:G2-ZP}
\end{align}
where the current experimental result suggests  $\Delta a_\mu= (26.1\pm8.0)\times 10^{-10}$~\cite{Hagiwara:2011af}.
\footnote{$Z$ boson decays are also important if sizable muon $g-2$ is expected~\cite{Nomura:2019jxj}.}


\subsection{Dark matter candidate}
{DM candidate in the model is the lightest neutral extra fermion that is denoted as $\psi_{R_1} \equiv X_R$ in the following. 
In our construction, DM may decay via $X \nu \chi $ coupling since $\chi$ can transferred to $\varphi_{A}$s through interactions in the potential associated with coupling $\lambda_\chi$ and $\tilde \lambda_\chi$.
To make DM lifetime sufficiently long, we here assume $\lambda_{\chi}$ and $\tilde \lambda_{\chi}$ are very small and/or scalar bosons associated with $\varphi_A$ are very heavy. 
Then decay width of DM can be very small so that DM is stable in cosmological time scale. 
}

{\it Relic density of Dark Matter}:
The relic density is severely restricted by the current experiment $\Omega h^2\approx0.12$~\cite{pdg}. 
In order to estimate the relic density, one has to know the interactions between DMs where the formulation is found as follows~\cite{DEramo:2010keq}:
\begin{align}
&\langle\sigma v_{\rm rel}\rangle_i=\sum_{i}\frac{\int_{4M^2_X}^\infty ds\sqrt{s-4M_X^2} W_{XX}^i K_1\left(\frac{\sqrt s}{M_X}x\right)}{16M_X^5x^{-1}K_2(x)^2},\\
&W_{XX}^i=\frac{1}{32\pi^2}\sqrt{1-\frac{4M_X^2}{s}}\int d\Omega |\bar {\cal M}_i|^2,
\end{align}

\begin{align}
&\Omega h^2\approx \frac{1.07\times 10^9 {\rm GeV}^{-1}}{\sqrt{g_*(x_f)} M_{pl} J_i(x_f)},\quad
J_i(x_f)=\sum_i \int_{x_f}^\infty \frac{\langle\sigma v_{\rm rel}\rangle_i}{x^2},
\end{align}
where $g_*(x_F)\approx 100$ is the total number of effective relativistic degrees of freedom at the time of freeze-out,
$M_{\rm pl}=1.22\times 10^{19}[{\rm GeV}] $ is Planck mass. 
$x_f$ is also found as follows~\cite{DEramo:2010keq}:
\begin{align}
x_f\simeq \ln\left[0.038c(c+2)\langle\sigma v_{\rm anni}\rangle\frac{g M_X M_{pl}}{\sqrt{g_*x_f}}\right]
+\ln\left[1+\frac{c+1}{c+2}\frac{\langle\sigma v_{\rm semi-anni}\rangle}{\langle\sigma v_{\rm anni}\rangle}\right],\label{eq:xf}
\end{align}
where $c=\sqrt2-1$, $\langle\sigma v_{\rm anni}\rangle$ is the total cross section for the DM annihilation modes,
while $\langle\sigma v_{\rm semi-anni}\rangle$ is the total cross section for the DM {\it semi}-annihilation modes.
The second term in Eq.~(\ref{eq:xf}) reflects on the effect of the semi-annihilation processes.\\
In our case, the valid Lagrangian in terms of mass eigenstate is given by
\begin{align}
-{\cal L}&=
Y_{i1} \bar\nu_i P_R X\left[(c_RH_1+s_R H_2)+i(c_IA_1+s_I A_2)\right] \nn\\
& +
(Y_\Delta^R)_{11} \bar X P_R X
\left[(-s_RH_1+c_R H_2)+i(-s_IA_1+c_I A_2)\right]   +{\rm h.c.},\\
& 
 (Y_\Delta^R)_{11}\equiv \sum_{a,b=1}^3\Omega^T_{1,a}g_{R_{ab}}\Omega^*_{b,1},
\end{align}
where $(Y_\Delta^R)_{11}=g_{R_{11}}/2$ in the case in which $g_R$ is diagonal matrix. 
Next task is to compute each of the mass invariant square $|{\cal \bar M}_i|^2$, and we define each of modes as follows:
$|{\cal \bar M}_1|^2$ is the process of $X\bar X\to \nu_i\bar\nu_j$, 
$|{\cal\bar M}_2|^2$ is $X X\to  \nu_i\nu_j$, 
$|{\cal\bar M}_3|^2$ is  $X \bar X\to X\bar\nu_i(\nu_i\bar X)$, 
Here $|{\cal \bar M}_{1,2}|^2$ correspond to $\langle\sigma v_{\rm anni}\rangle$, while 
$|{\cal \bar M}_{3}|^2$ corresponds to $\langle\sigma v_{\rm semi-anni}\rangle$.
Then, each of invariant mass square is give by
\begin{align}
|{\cal\bar M}_1|^2&=\sum_{i,j=1-3}|Y_{i1}Y^\dag_{1j}|^2
\left(\left|\sum_{J=H_{1,2}}^{A_{1,2}}S_J T_J\right|^2(p_1\cdot k_1)(p_2\cdot k_2)
+\left|\sum_{J=H_{1,2}}^{A_{1,2}}S_J U_J\right|^2(p_1\cdot k_2)(p_2\cdot k_1)\right.\nn\\
&\left.\hspace{3cm}
-{\rm Re}\left[\sum_{J=H_{1,2}}^{A_{1,2}}S_J^2 T_J U_J^*\right]M_X^2(k_1\cdot k_2)\right),\\
|{\cal\bar M}_2|^2&=\sum_{i,j=1-3} |Y_{i1}Y^\dag_{1j}|^2
\left(\left|\sum_{J=H_{1,2}}^{A_{1,2}}S_J T_J\right|^2(p_1\cdot k_1)(p_2\cdot k_2)
+\left|\sum_{J=H_{1,2}}^{A_{1,2}}S_J U_J\right|^2(p_1\cdot k_2)(p_2\cdot k_1)\right.\nn\\
&\left.\hspace{1cm}
-{\rm Re}\left[\sum_{J=H_{1,2}}^{A_{1,2}}S_J^2 T_J U_J^*\right](2p_1\cdot k_1 p_2\cdot k_2- p_1\cdot p_2 k_1\cdot k_2)\right),\\
|{\cal\bar M}_3|^2&=2 \sum_{i=1-3} Y_{i1}Y^\dag_{1i}|(Y^R_\Delta)_{11}|^2(p_1\cdot p_2)(k_1\cdot k_2)\nn\\
&\times
\left|\frac{s_Rc_R}{s-m^2_{H_1}+im_{H_1}\Gamma_{H_1}}-\frac{s_Rc_R}{s-m^2_{H_2}+im_{H_2}\Gamma_{H_2}}
+
\frac{s_Ic_I}{s-m^2_{A_1}+im_{A_1}\Gamma_{A_1}}-\frac{s_Ic_I}{s-m^2_{A_2}+im_{A_2}\Gamma_{A_2}}
\right|^2,
\end{align}
where we define $T_{J}\equiv 1/(M_X^2-m_{J}^2-2p_1\cdot k_1)$, $U_J\equiv 1/(M_X^2+m_{J}^2-2p_1\cdot k_2)$.

\subsection{Numerical analysis and phenomenology}

Now that all of formulae are provided, we will move on to the global numerical analysis.
First of all, we fix the regions of our input parameters as:
\begin{align}
&O_{ij}\in [-i,\pi+10i],\ g_R\in [0.1,\sqrt{4\pi}], \ s_{R/I}\in [-1,1],\ M_L\in [100,1000]\ {\rm GeV},\\
& \mu_{R/L}\in [10,100]\ {\rm GeV},\  (m_{H_{1,2}},m_{A_{1,2}}) \in [100,5000]\ {\rm GeV},
\  (\Gamma_{H_{1,2}},\Gamma_{A_{1,2}}) \in [0,10]\ {\rm GeV},\nn
\end{align}
where $O_{ij}$ (i=1-3, j=1-6) consists of six independent parameters.
Then we search the allowed region to satisfy the neutrino oscillation data, LFVs, and relic density of DM candidate.

\begin{figure}[t]
\begin{center}
\includegraphics[width=10cm]{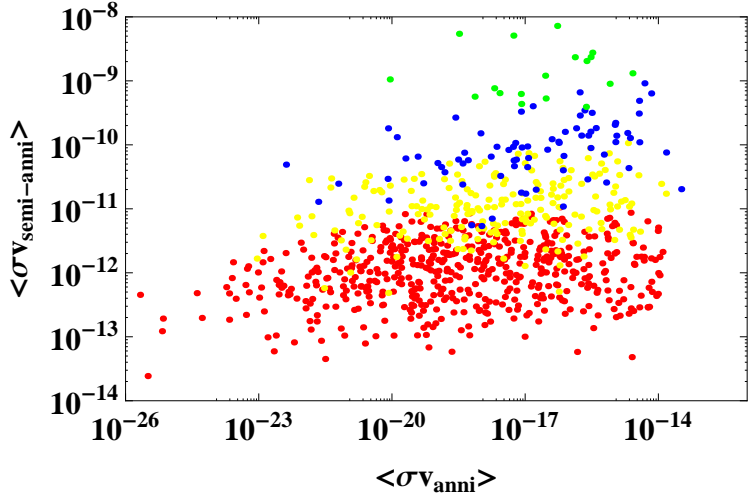}
\caption{Cross sections between annihilation and semi-annihilation processes, where green region demonstrates $0<\Omega h^2\le1$, blue one $1<\Omega h^2\le10$, yellow one $10<\Omega h^2\le100$, and red one $100<\Omega h^2\le1000$.
 }   \label{fig:cross}
\end{center}\end{figure}

Fig.~\ref{fig:cross} represents the cross sections between annihilation and semi-annihilation processes, where green region demonstrates $0<\Omega h^2\le1$, blue one $1<\Omega h^2\le10$, yellow one $10<\Omega h^2\le100$, and red one $100<\Omega h^2\le1000$. Since observed relic density is about 0.1, the green region is favored. Threrefore, semi-annihilation processes are crucial
in our scenario.

\begin{figure}[t]
\begin{center}
\includegraphics[width=10cm]{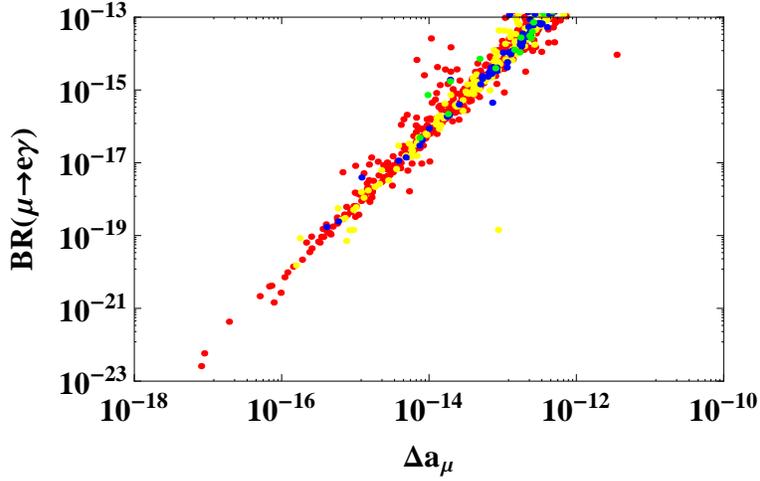}
\caption{Branching ratio of $\mu\to e\gamma$ in terms of muon $g-2$, where green region demonstarates $0<\Omega h^2\le1$, blue one $1<\Omega h^2\le10$, yellow one $10<\Omega h^2\le100$, and red one $100<\Omega h^2\le1000$. }   \label{fig:am-lfvs}
\end{center}\end{figure}

Fig.~\ref{fig:am-lfvs} show branching ratio of $\mu\to e\gamma$ in terms of muon $g-2$, where each of colored region gives the same meaning as Fig.~\ref{fig:cross}. It suggests that typical size of muon $g-2$ is $10^{-14}\sim 10^{-12}$, which is smaller than the observed result by two order of magnitude.~\footnote{If we have fine-tunings for our parameters; e.g., resonant points for semi-annhilation processes, then one might be able to find sizable muon $g-2$. But this is beyond our scope.}

\section{Summary and discussions}
We have proposed  radiative neutrino mass model with gauged hidden $U(1)$ symmetry,
and discussed fermionic DM candidate that has semi-annihilation processes in the relic density, LFVs taking into account experimental constraints, and muon $g-2$.
{In the model, Majorana mass term of extra neutral fermions is generated at one-loop level.
Then active neutrino mass is generated at one loop level which contains contribution from the Majorana mass of extra neutral fermions.
Thus our active neutrino mass generation is at two loop level in total.
The lightest extra neutral fermion can be a DM candidate and its stability is obtained choosing inert singlet scalar mixing to be very small.
Then our DM induces semi-annihilation processes. 
}

We have demonstrated global analysis to investigate if our model can satisfy neutrino oscillation data, lepton flavor violations, and relic density of dark matter candidate, 
and shown that semi-annihilation modes play a crucial role in finding observed relic density.
Typical muon $g-2$ is found in the region of the order of $10^{-14}\sim 10^{-12}$, which is smaller than the experimental result by two order of magnitude.

\section*{Acknowledgments}
This research is supported by the Ministry of Science, ICT and Future Planning, Gyeongsangbuk-do and Pohang City (H.O.). 
H. O. is sincerely grateful for KIAS and all the members.

\end{document}